\documentclass[twocolumn]{aastex62}
\usepackage{natbib}
\usepackage{multirow}
\usepackage{amsmath}

\begin{document}

\title{Brightest Cluster Galaxy Alignments in Merging Clusters}

\shorttitle{Brightest Cluster Galaxy Alignments in Merging Clusters}

\author[0000-0002-0813-5888]{David Wittman}
\affiliation{Physics Department, University of California, Davis, CA 95616}
\author{Drake Foote}
\affiliation{Physics Department, University of California, Davis, CA 95616}
\author{Nathan Golovich}
\affiliation{Lawrence Livermore National Laboratory, 7000 East
  Avenue, Livermore, CA 94550, USA}
\correspondingauthor{David Wittman}
\email{dwittman@physics.ucdavis.edu}

\begin{abstract}
The orientations of brightest cluster galaxies (BCGs) and their host clusters tend to be aligned, but the mechanism driving this is not clear. To probe the role of cluster mergers in this process, we quantify alignments of 38 BCGs in 22 clusters undergoing major mergers (up to $\sim 1$ Gyr after first pericenter). We find alignments entirely consistent with those of clusters in general. This suggests that alignments are robust against major cluster mergers. If, conversely, major cluster mergers actually help orient the BCG, such a process is acting quickly because the orientation is in place within $\sim 1$ Gyr after first pericenter.
\end{abstract}


\section{Introduction}
Galaxies are not randomly oriented, and galaxy alignments are a topic of increasing interest; see \citet{Joachimi15} for a recent review. We focus here on the specific case of brightest cluster galaxies (BCGs) and their alignment with the clusters in which they are embedded \citep{Binggeli82}.  In the modern era of large surveys, \citet{Niederste-Ostholt10} conducted a seminal study on this topic by examining $\sim$10,000 clusters selected from the Sloan Digital Sky Survey, measuring major axes of both BCGs and clusters using visible-wavelength light. They confirmed at high significance that the major axes tend to be aligned. Second-brightest cluster galaxies were much less likely to be aligned, and third-brightest galaxies were random, suggesting that BCGs uniquely undergo some alignment process. This is further supported by their finding that more dominant BCGs exhibit stronger alignments. \citet{Huang16} updated and extended this type of study by considering alignment strength as a function of central and satellite galaxy size, luminosity, and color, while \citet{Donahue16} showed that the BCG-cluster alignment is preserved even when cluster shapes are measured by X-rays, Sunyaev-Sel'dovich effect, and/or gravitational lensing. 

Theoretical treatments of BCG alignments have focused on anisotropic infall along filaments, primordial alignments in protoclusters and galaxies, and gradual alignment to the local field through gravitational torques \citep{West1994,Catelan1996,Libeskind2013}.  To get a more direct handle on how and when these alignments formed, \citet{West17BCG} examined a sample of higher-redshift clusters, $0.19<z<1.8$. By examining the alignment of BCGs with the red sequence galaxy population, they found no redshift trend: BCG-cluster alignments persist to a lookback time of $\sim$ 10 Gyr, along with a lack of alignment of other cluster members.  

If BCG alignments depend little on cosmic era, perhaps they depend on events in the life of each cluster, such as mergers or the slower accretion of matter along the  filaments of the cosmic web. This paper probes the role of mergers by quantifying BCG alignments in a sample of clusters observed after undergoing first pericenter in major binary mergers.  

The paper is organized as follows: in \S \ref{sec-sample}, we outline our sample of BCGs within our ensemble of merging galaxy clusters.  In \S \ref{sec-methods} we present the geometry and methods our of  analysis. In \S \ref{sec-results} we present the results of our study,  and we discuss our findings in \S \ref{sec-discussion}. 

\section{Sample}\label{sec-sample}

We adopt the Merging Cluster Collaboration (MCC) sample of 29 merging systems \citep{Golovich18a,Golovich18b}. These systems were selected to be post-pericenter based on their radio morphology, with radio relics tracing shocks in the cluster gas \citep{Ensslin1998}.  \citet{Golovich18a} found that most of these systems have bimodal (or multimodal) galaxy distributions on the sky, but unimodal line-of-sight velocity distributions. This suggests that the mergers are occurring largely in the plane of the sky. \citet{Golovich18b} found that the X-ray morphology and three dimensional (projected and line of sight velocity) galaxy distributions are also consistent with mergers occurring in the plane of the sky observed after first pericenter, but before second pericenter. Given that the time between pericenters is of order 1--2 Gyr, we expect observations of these systems to be typically $\sim$ 1 Gyr after first pericenter. Indeed, the dynamics of several clusters in our sample have been modeled with the MCMAC\footnote{\url{https://github.com/MCTwo/MCMAC}} code \citep{Dawson2012} with the time since pericenter ranging from $\sim$ 0.7 to 1.2 Gyr \citep{GolovichMACS1149,GolovichZwCl0008,vanWeeren2017Abell3411, BensonZwCl2341}.

The absence of systems that are merging along the line of sight allows us to probe the alignment between BCGs and the subcluster separation vector, which is presumably a proxy for the direction of the filament along which the clusters are merging, as well as the major axis of the eventual merged cluster. 

\citet{Golovich18b} used a Gaussian mixture model to group galaxy positions and redshifts into subclusters in each system, using the Bayesian information criterion to prevent oversplitting. They checked the mean redshift of each subcluster to identify subclusters that could be physically associated.  If this yielded two physically associated subclusters, \citet{Golovich18b} checked that the radio and X-ray morphologies were consistent with a recent pericenter passage of the two subclusters: X-rays are seen between the subclusters and extended along the subcluster separation vector, which when extended beyond the subclusters bisects the radio relic(s).  For systems with more than two subclusters, they used this geometric model to identify the two subclusters most likely to have created the radio and X-ray morphologies with a recent pericenter passage.  This was not possible in a few systems, which we have eliminated from our analysis in this paper. This was due to the lack of Subaru/SuprimeCam imaging (Abell 2443 and PSZ1 G108.18-1153) or an unclear merger scenario (Abell 746, Abell 2255, Abell 2345, Abell 2744, PLCK G287.0+32.9).  Eliminating these systems leaves 22 systems with subcluster pairs tied to the X-ray and radio morphologies; these are also typically the two most massive subclusters in a system, judging by the relative sizes of their velocity dispersions. Some of these 22 systems have additional subclusters not clearly tied to the X-ray and radio emission; those subclusters could be infalling for the first time, or too low-mass to generate substantial X-ray and radio effects.  This paper uses the BCGs of the two subclusters most closely associated with the X-ray and radio emission in each system.

From the Subaru/SuprimeCam images \citep{Golovich18a} of these 22 systems, we prepare 150$\times$150 kpc$^{2}$ cutout images\footnote{Assuming a flat $\Lambda$CDM universe with $H_0 = 70$ km s$^{-1}$ Mpc$^{-1}$ and $\Omega_m=0.3$.} of the 44 BCGs assigned to the merging subclusters.  The cluster-wide BCG cutouts were presented in \citet{Golovich18b} for each system. All analysis is completed on these cutouts as described in the following section. For each cluster we used the deepest image listed in Table 2 of \citet{Golovich18a} (usually $R$-band, with about 20\% $I$ band).  We refer readers to that paper for details regarding the observations and data reduction.

\section{Methods}\label{sec-methods}

We used GALFIT \citep{GALFIT2010}, a 2D surface brightness fitting software package, to fit a S\'{e}rsic profile to each BCG. The S\'{e}rsic model parameters include the central coordinates, integrated magnitude, half-light radius, S\'{e}rsic index, axis ratio, and major axis position angle (PA)---which is the parameter of interest here.  We fit to the entire 150$\times$150 kpc$^{2}$ cutout image of each BCG as seen in \citet{Golovich18a}.  We used GALFIT without a point-spread function (PSF) model because our BCGs are overwhelmingly larger than the PSF. We did make extensive use of pixel masking as described in the next paragraph.

\begin{figure*}
\centerline{\includegraphics[width=3.5in]{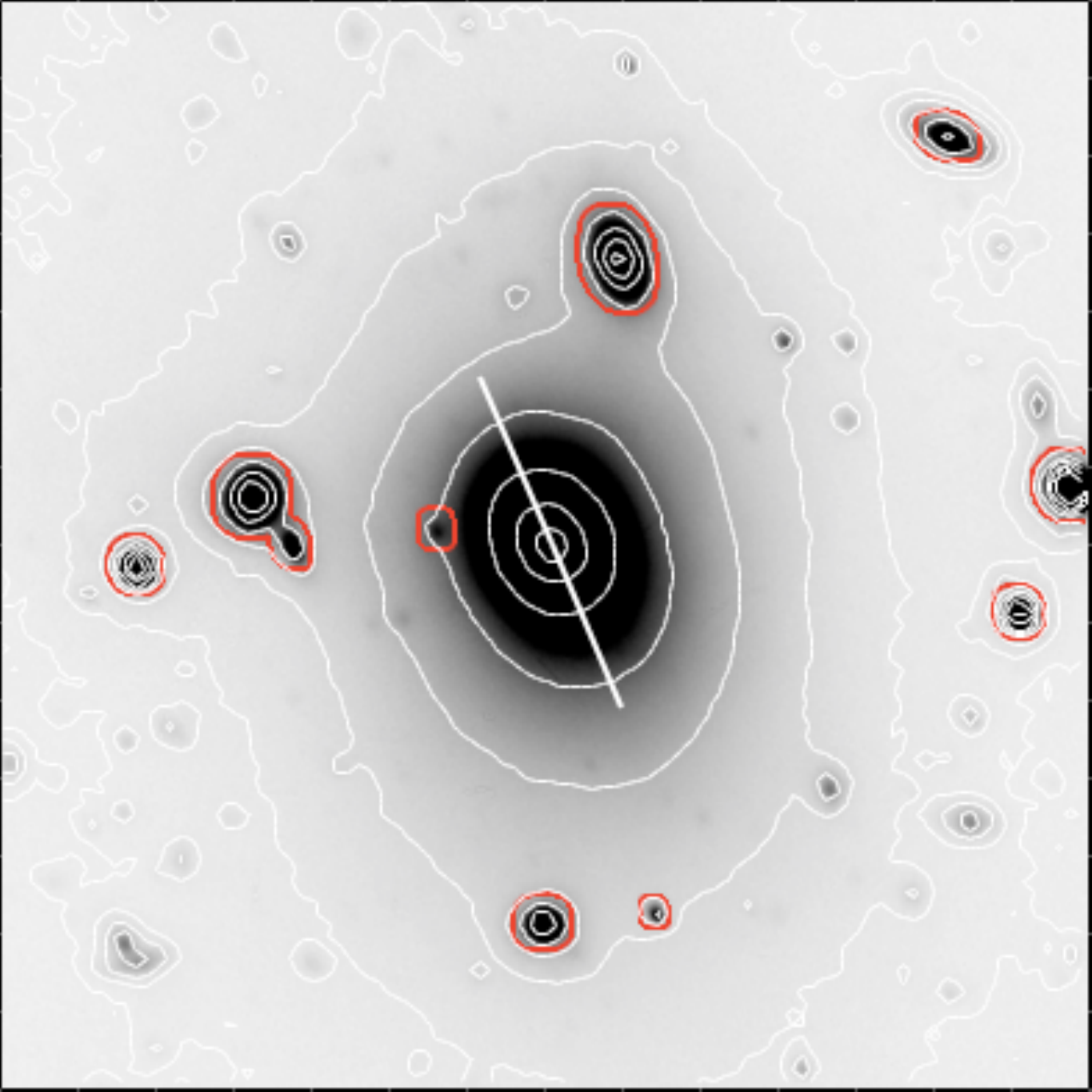}\hskip2mm\includegraphics[width=3.5in]{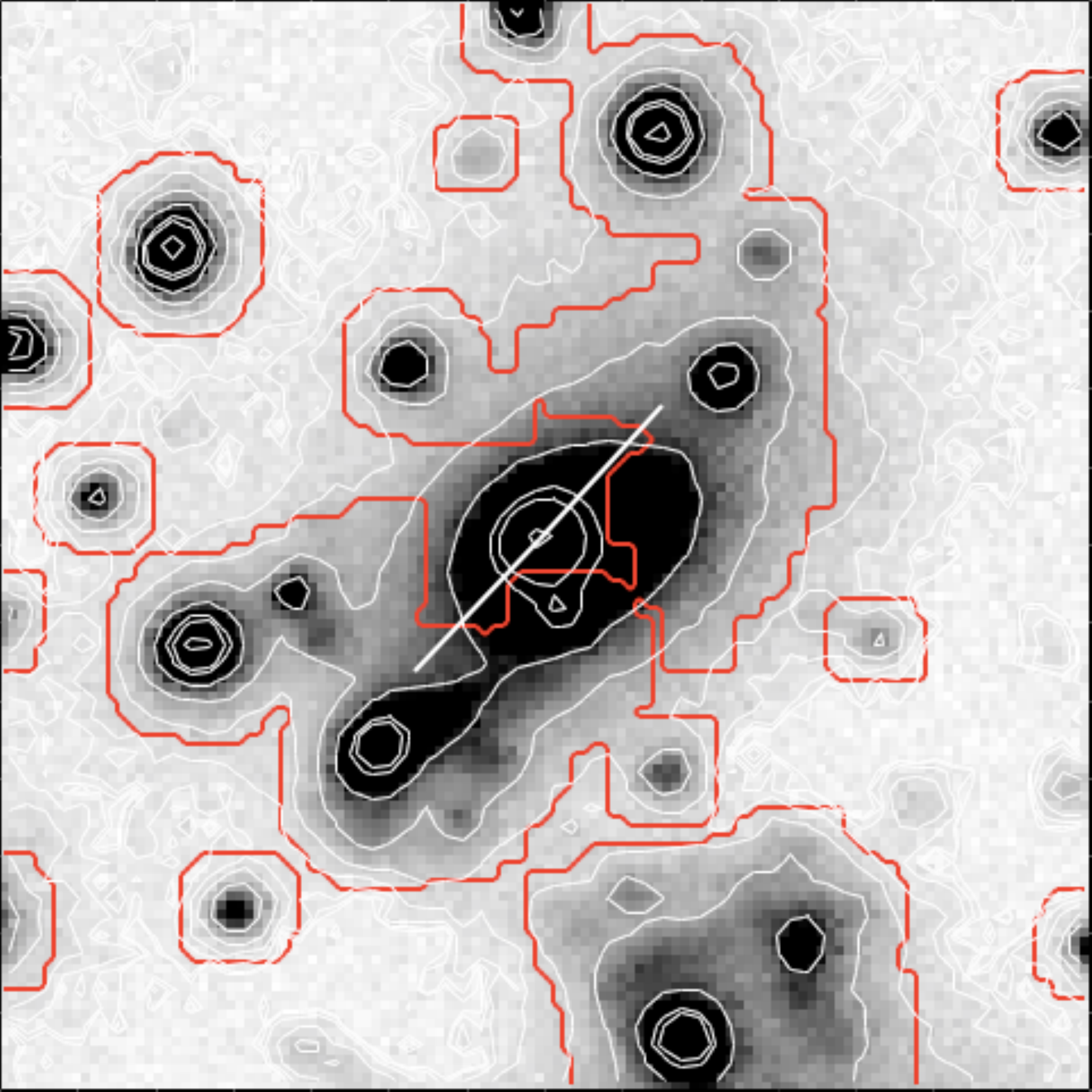}}
\caption{A BCG in a typical field (Abell 523 North, left) and in the most crowded field (MACS J1149.5+2223 North, right). Bars denote the BCG major axis PA found by GALFIT, red contours show the masked regions, and white contours show the surface brightness spaced by factors of 2.5. An extra contour has been added to the right-hand panel to highlight the presence of a near neighbor that was masked. Each image is 150 kpc across, and each bar is 50 kpc across. }\label{fig-masks}
\end{figure*}

The pixel mask aimed to remove various foreground and background objects such as stars or other galaxies from the image. We created a custom Python script to identify all pixels meeting both of two conditions: (i) intensity greater than five percent of the peak brightness of the BCG; and (ii) having a local intensity minimum along the vector between the pixel and the BCG peak. The red contours in Figure~\ref{fig-masks} show the masks for a typical field, Abell 523 North, and the most crowded field, MACS J1149.5+2223 North. For four BCGs (Abell 2034 South, Abell 2061 North, Abell 2061 South, and RXC J1314.4-2515 East) this algorithm masked most of the pixels in the BCG itself, so we did not attempt to fit the data.

GALFIT generally converges to a good fit regardless of the initial guess, so we used simple metrics for the initial guesses. We initialized the profile centroid to the brightest pixel location, the half-light radius to the half width at half-maximum (HWHM), and the axis ratio and PA based on visual inspection. We started the S\'{e}rsic index at 4.0, and restarted at 2.0 if GALFIT did not converge starting from 4.0.

GALFIT reliably found models that agreed with manual inspection (see the arrows in Figure~\ref{fig-masks}). Given the complicated environments, however, the uncertainties in the fit parameters are likely to be larger than computed by GALFIT's simple photon noise model. Therefore, we tested the robustness of the results in different ways:
\begin{itemize}\item First, we tested the effect of varying the masking procedure by fitting each BCG with no mask at all. In a few cases omitting the mask made a substantial difference, and the masked fit was clearly superior to the unmasked fit.  But in many cases the masking variation changed the result by only 1--2$^\circ$, presumably due to the BCG's dominance and the nearly isotropic distribution of neighbors.  Thus, masking even fainter neighbors---whose distribution is even more isotropic---is likely to have an even smaller effect. We conclude that the masking is sufficient to determine the PA to $\sim 1^\circ$.  

\item Second, we assessed the dependence of the GALFIT result on its initial parameters. We reran GALFIT 121 times for each BCG with a grid of 11 initial guesses for the axis ratio and for the PA, spanning $\pm0.1$ in axis ratio and $\pm5^\circ$ in PA about the fiducial initial guess.  We then computed the rms variation of PA over these 121 samples. We dropped two BCGs (ZwCl 1856.8+6616 South and ZwCl 2341.1+0000 North) from the sample due to large rms variation ($> 7^\circ$).  Of the remaining 38 BCGs, the median (mean) rms variation was 0.3$^\circ$ (1.3$^\circ$).  

\item Third, we tested the internal consistency of each BCG by refitting to 50 kpc and 100 kpc diameter postage stamps and tabulating the most extreme absolute value of the three PA differences PA$_{\rm 50 kpc}-$PA$_{\rm 100 kpc}$, PA$_{\rm 100 kpc}-$PA$_{\rm 150 kpc}$, and PA$_{\rm 50 kpc}-$PA$_{\rm 150 kpc}$.  The median (mean) value of this statistic across 38 BCGs was 7$^\circ$ (10$^\circ$). In many cases there was a monotonic PA trend from 50 to 150 kpc indicating isophotal twist. BCG alignment with the merger axis could thus in principle be a function of radius within the BCG.  However, any such trend in our sample will be subtle because the isophotal twist from 50--150 kpc is typically only $\approx 10^\circ$. Therefore, we focus here on the PA within a single well-defined diameter, 150 kpc, that captures most of the light.  
\end{itemize}
In summary, these robustness tests caused us to drop two more BCGs due to sensitivity to the initial fit parameters.  Four had already been dropped due to paucity of unmasked pixels, leaving 38 BGCs with a measurement uncertainty that depends on crowding but is typically 1--2$^\circ$.  While each BCG major axis PA may vary by $\approx 10^\circ$ as a function of radius, we will consistently quote the PA of the light within a 150 kpc diameter.

\begin{table*}
\begin{center}
\begin{tabular}{crrccrc}\multicolumn{2}{c}{\underline{\hskip1.5cm Merger\hskip1.5cm}}& \multicolumn{5}{c}{\underline{\hskip3.4cm BCG \hskip2.8cm}}\\ Name&PA$_{\rm m}$&ID\tablenotemark{a}&RA (J2000)&Dec&PA$_{\rm BCG}$&Axis ratio\\ \hline
\multirow{2}{*}{1RXS J0603.3+4212}&\multirow{2}{*}{164}&N&06:03:16.7&+42:14:41&48&0.88\\
&&S&06:03:24.3&+42:09:31&173&0.82\\ \hline
\multirow{2}{*}{Abell 115}&\multirow{2}{*}{152}&N&00:55:50.6&+26:24:37&148&0.57\\
&&S&00:56:00.3&+26:20:33&14&0.88\\ \hline
\multirow{2}{*}{Abell 521}&\multirow{2}{*}{149}&N&04:54:06.9&$-$10:13:25&136&0.77\\
&&S&04:54:15.4&$-$10:16:14&136&0.78\\ \hline
\multirow{2}{*}{Abell 523}&\multirow{2}{*}{14}&N&04:59:13.0&+08:49:42&22&0.72\\
&&S&04:59:06.6&+08:43:49&2&0.80\\ \hline
\multirow{2}{*}{Abell 781}&\multirow{2}{*}{165}&N&09:20:22.4&+30:32:30&0&0.72\\
&&S&09:20:25.7&+30:29:38&27&0.79\\ \hline
\multirow{2}{*}{Abell 1240}&\multirow{2}{*}{175}&N&11:23:35.3&+43:08:31&16&0.66\\
&&S&11:23:37.7&+43:03:28&9&0.78\\ \hline
\multirow{2}{*}{Abell 1300}&\multirow{2}{*}{55}&E&11:32:00.3&$-$19:54:41&174&0.69\\
&&W&11:31:54.2&$-$19:55:40&131&0.59\\ \hline
\multirow{2}{*}{Abell 1612}&\multirow{2}{*}{120}&N&12:47:33.3&$-$02:47:12&164&0.70\\
&&S&12:47:51.4&$-$02:49:50&132&0.72\\ \hline
Abell 2034&176&N&15:10:10.2&+33:34:01&95&0.82\\ \hline
\multirow{2}{*}{Abell 2163}&\multirow{2}{*}{81}&E&16:15:49.0&$-$06:08:42&91&0.71\\
&&W&16:15:33.5&$-$06:09:17&78&0.66\\ \hline
\multirow{2}{*}{Abell 3365}&\multirow{2}{*}{74}&E&05:48:43.2&$-$21:54:56&69&0.71\\
&&W&05:48:18.7&$-$21:56:31&88&0.68\\ \hline
\multirow{2}{*}{Abell 3411}&\multirow{2}{*}{152}&N&08:41:52.9&$-$17:28:05&13&0.75\\
&&S&08:42:07.1&$-$17:34:41&121&0.71\\ \hline
\multirow{2}{*}{CIZA J2242.8+5301}&\multirow{2}{*}{8}&N&22:42:52.6&+53:04:51&127&0.88\\
&&S&22:42:40.8&+52:58:54&25&0.80\\ \hline
\multirow{2}{*}{MACS J1149.5+2223}&\multirow{2}{*}{137}&N&11:49:35.7&+22:23:55&136&0.72\\
&&S&11:49:43.0&+22:22:07&128&0.82\\ \hline
\multirow{2}{*}{MACS J1752.0+4440}&\multirow{2}{*}{61}&E&17:52:11.9&+44:41:01&97&0.71\\
&&W&17:51:53.4&+44:39:14&60&0.67\\ \hline
\multirow{2}{*}{RXC J1053.7+5452}&\multirow{2}{*}{121}&E&10:54:11.2&+54:50:18&67&0.89\\
&&W&10:53:36.5&+54:52:04&133&0.91\\ \hline
RXC J1314.4-2515&82&W&13:14:22.1&$-$25:15:46&55&0.70\\ \hline
\multirow{2}{*}{ZwCl 0008.8+5215}&\multirow{2}{*}{77}&E&00:12:18.9&+52:33:47&62&0.78\\
&&W&00:11:21.8&+52:31:45&81&0.87\\ \hline
\multirow{2}{*}{ZwCl 1447.2+2619}&\multirow{2}{*}{179}&N&14:49:31.1&+26:08:37&13&0.75\\
&&S&14:49:29.7&+26:02:55&35&0.73\\ \hline
ZwCl 1856.8+6616&171&N&18:56:33.7&+66:25:16&106&0.82\\ \hline
ZwCl 2341.1+0000&146&S&23:43:47.5&+00:15:24&146&0.84\\ \hline
\end{tabular}
\tablenotetext{a}{Identifier: (N)orth, (S)outh, (E)ast, (W)est}
\end{center}
\caption{Merger and BCG parameters. Position angle (PA) is measured in degrees from north toward east.}\label{tab1}\end{table*}

\begin{figure} \includegraphics[width=3.5in]{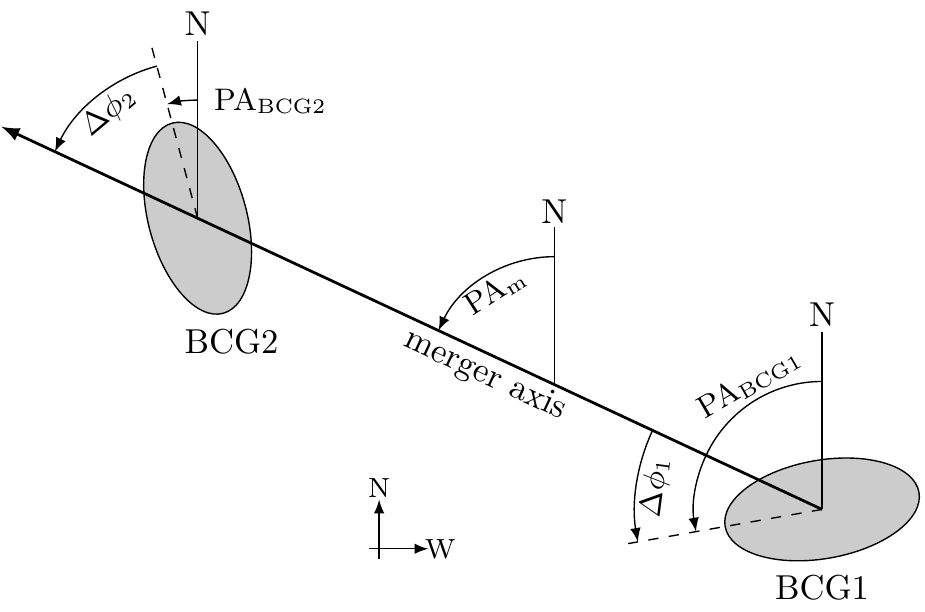} \caption{Schematic representation of the geometry for a bimodal merger showing the position angles for each BCG and for the merger axis (the line connecting the BCGs). PAs are defined as increasing from north to east.  The BCG-merger axis angle $\Delta\phi\equiv|{\rm PA}_{\rm BCG}-{\rm PA}_{\rm m}|$ is defined only on the interval $[0^\circ,90^\circ]$ to reflect invariance to rotation by 180$^\circ$ and/or reflection across the merger axis.}  \label{fig-geometry} \end{figure}

Table~\ref{tab1} lists the fit results for each BCG. PA$_{\rm BCG}$ is defined as the angle between north (increasing to the east) and the BCG major axis as shown in Figure~\ref{fig-geometry}. The merger axis PA$_{\rm m}$ is similarly defined using the line connecting the two BCGs. Then, for each BCG we compute $\Delta\phi\equiv |\rm{PA}_{\rm BCG}-\rm{PA}_{\rm m}|$.  Finally, recognizing that $\Delta\phi$ and $180^\circ -\Delta\phi$ represent the same physical situation, we fold $\Delta\phi$ into the range $[0^\circ,90^\circ]$. For example, PA$_{\rm BCG}=85^\circ$ and PA$_{\rm merger}=-80^\circ$ represent a PA difference of only $15^\circ$ after folding.

\section{Results}\label{sec-results}

\begin{figure}
\includegraphics[width=\columnwidth]{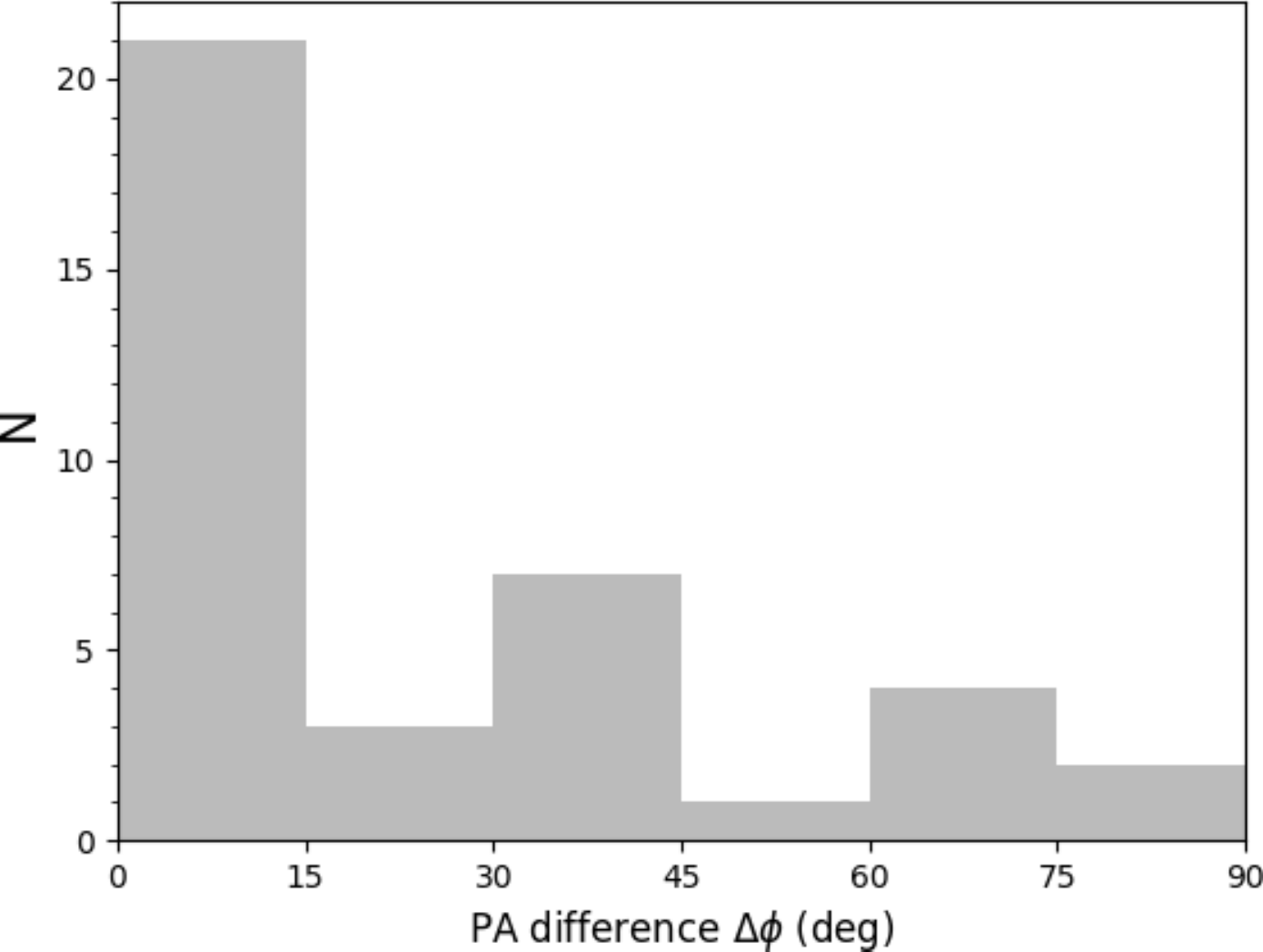}
\caption{Histogram of $\Delta
\phi$ values for the 38 BCGs in Table~\ref{tab1}. The excess of low $\Delta \phi$ values indicates, with high significance, that BCGs tend to be oriented along the merger axis.}\label{fig-dphihist}
\end{figure}

Figure~\ref{fig-dphihist} shows the histogram of $\Delta \phi$ values. For comparison, randomly oriented galaxies would yield a flat distribution in this figure. The observed distribution is significantly biased toward small $\Delta\phi$, i.e., BCGs tend to be aligned with the merger axis. According to a Kolmogorov-Smirnov test, this distribution is highly inconsistent with uniform ($p=1.2\times10^{-5}$).

\begin{figure}
\includegraphics[width=\columnwidth]{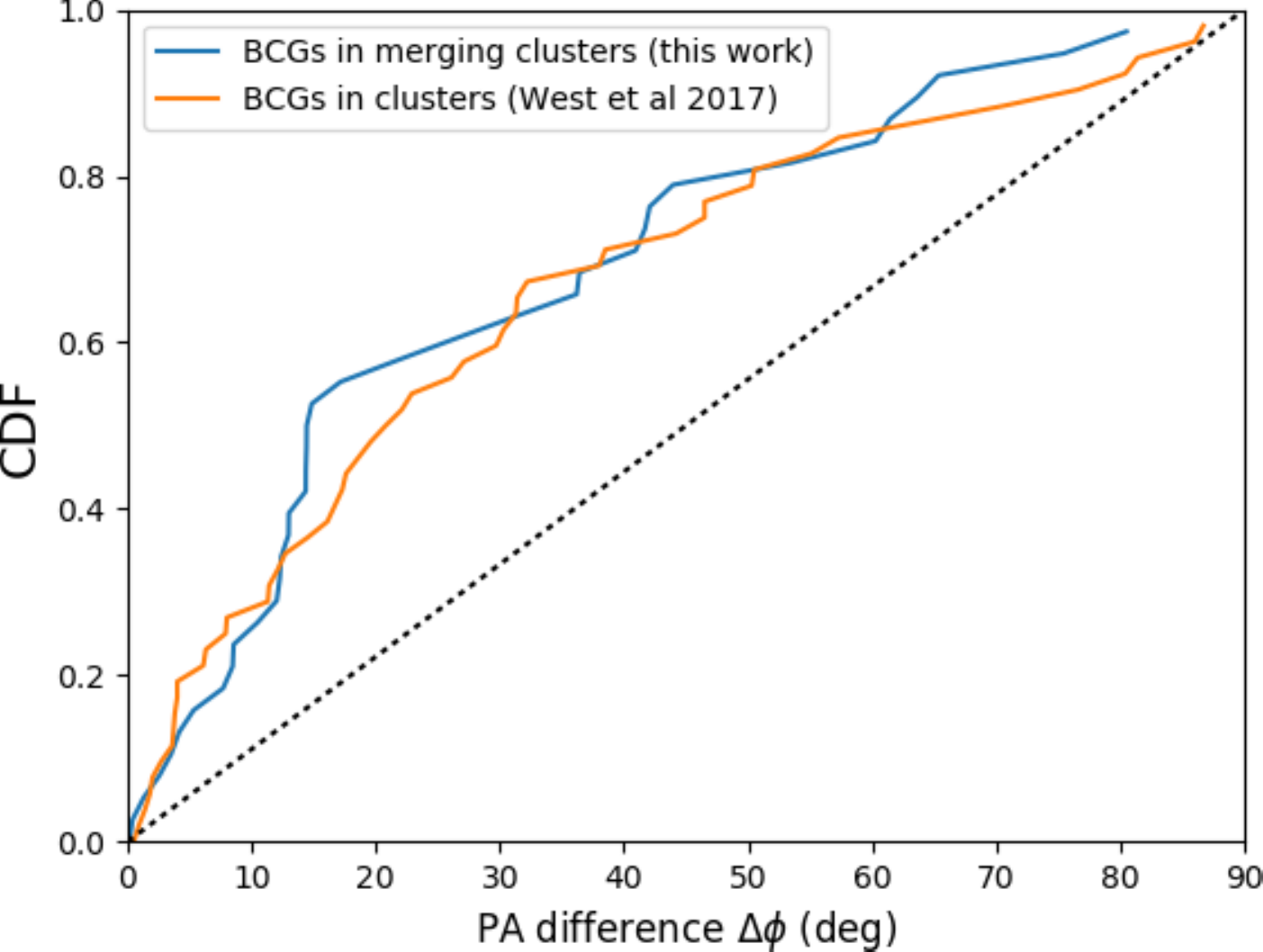}
\caption{Cumulative distribution functions for $\Delta\phi$ for our merging cluster sample and for the \citet{West17BCG} sample of more typical clusters.  The two distributions are consistent with each other, and each is inconsistent with a uniform distribution (dashed line). This suggests that mergers either preserve pre-existing alignments, or act to induce alignments on a timescale of $\lesssim 1$ Gyr.}\label{fig-CDF}
\end{figure}

Next, we compare the alignments in our sample with the more typical cluster sample of \citet{West17BCG}. Figure~\ref{fig-CDF} shows the cumulative distribution function (CDF) for each sample. The two CDFs are remarkably similar to each other: a two-sided Kolmogorov-Smirnov test indicates that the two distributions are consistent, $p=0.53$. The more sensitive Anderson-Darling test \citep{2012msma.book.....F} also finds consistency, $p>0.25$. This consistency suggests that mergers either preserve pre-existing alignments, or act to induce alignments on a timescale of $\lesssim 1$ Gyr.

\begin{figure}
\includegraphics[width=\columnwidth]{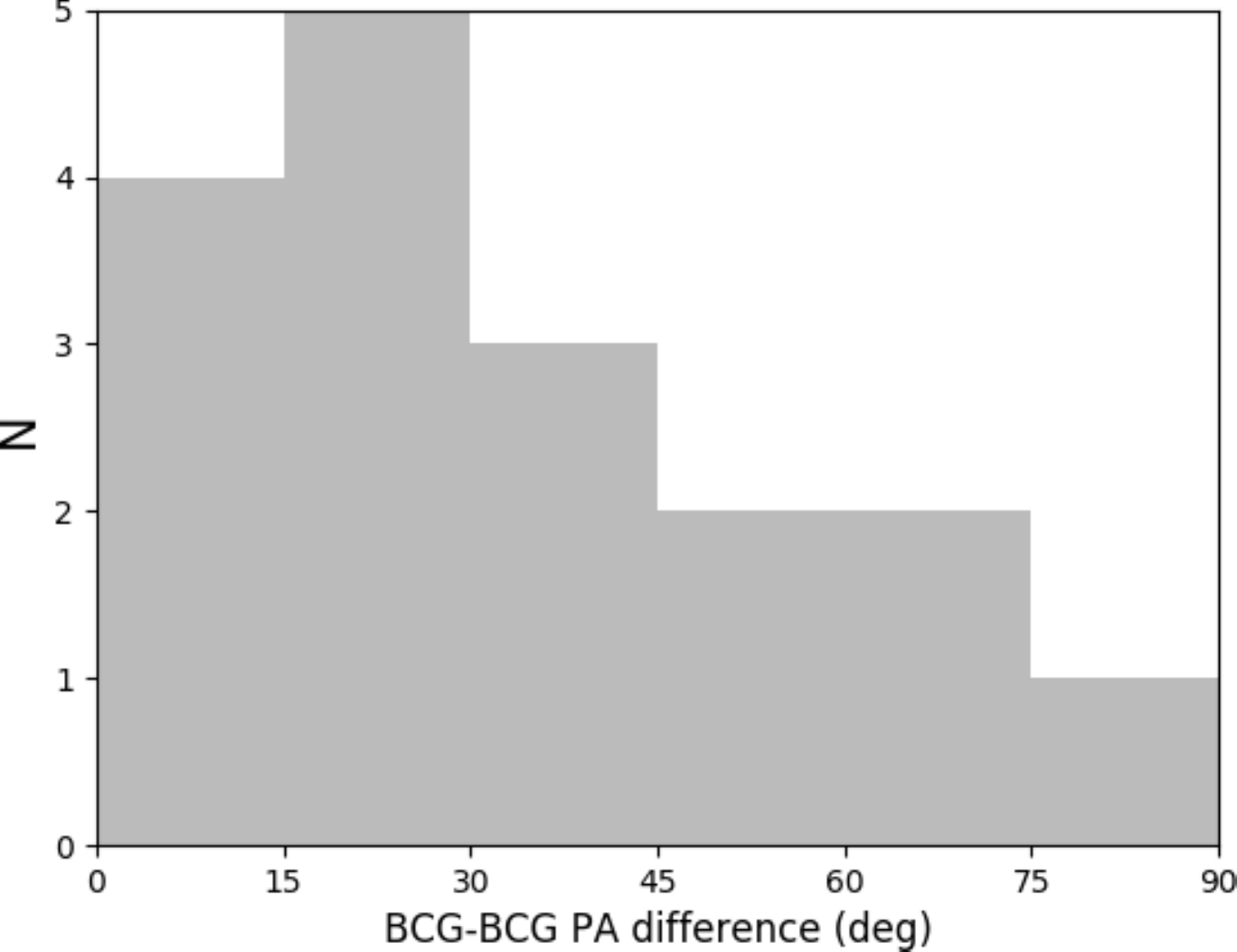}
\caption{Histogram of position angle differences for the 17 pairs of BCGs in Table~\ref{tab1}, indicating that BCGs correlate with each other but not as well as with the merger axis. As with Figure~\ref{fig-dphihist}, bins are 10$^\circ$ wide, but the numbers are cut by more than half due to missing BCG measurements in some systems.}\label{fig-BCGBCG}
\end{figure}

So far we have shown that each BCG PA is correlated with the merger axis. It could be that the BCGs directly correlate with each other and only less so with the merger axis, thus rendering the merger axis a mere confounding variable. We tested this by tabulating, for each merger with two well-measured BCGs, the angle between the major axes of the two BCGs. Figure~\ref{fig-BCGBCG} shows the resulting histogram, which is flatter than the histogram in Figure~\ref{fig-dphihist}. The relative weakness of the BCG-BCG correlation 
suggests that the separation vector is playing a central role in guiding the orientation of each BCG, rather than the BCGs affecting each other directly. This does not, however, specify the underlying physical mechanism and in particular does not shed light on whether that mechanism predates the merger. For example, if a merger can be modeled as infall along mostly opposing filaments, each BCG could be aligned with its host filament before the merger. 

We also checked the $\Delta\phi$ distribution of the 22 well-measured BCGs that remain after excluding systems that \citet{Golovich18b} found to be multimodal.  The distribution is consistent with that of the full sample. We found the same result for a subset of fifteen BCGs residing in the ``gold'' subsample of eight bimodal systems that \citet{Golovich18b} found to have merger axes particularly well defined by radio relics.  This suggests that, even in the more complicated systems, identification of the merger axis is a subdominant source of uncertainty in this analysis. 

We also looked for an alignment trend with BCG axis ratio; that is, are more elongated BCGs more likely to be aligned with the merger axis?  We would expect this trend if the observed scatter in $\Delta\phi$ were mostly due to measurement uncertainty in the BCG PA, because rounder BCGs have more uncertain PA. We performed a Spearman rank correlation test between BCG axis ratio and $\Delta\phi$, and found a correlation coefficient of 0.15 with a p-value of $p=0.37$. This suggests that measurement uncertainties are subdominant, and that the BCG-merger axis scatter is intrinsic. This claim is further supported by the size of the BCG-merger axis scatter (tens of degrees) compared to the measurement uncertainty of the BCG major axis PA (1--2$^\circ$; see \S\ref{sec-methods}).  Note, however, that we have defined the merger axis according to the BCG separation vector.  While the uncertainty on the PA of the {\it current} BCG separation vector is negligible, it is likely that this PA changes over time as the BCGs are unlikely to follow exactly radial trajectories. \citet{Golovich18b} found that the scatter between the current BCG separation vector PA and radio relic PA is a few tens of degrees, which supports the idea that the current BCG separation vector does not completely define the merger geometry. Hence, a substantial fraction of our observed scatter may be due to the limitations of having a single snapshot of each merger. Simulations may be the best tool for further probing this question.

\section{Discussion}\label{sec-discussion}

Our primary result is that BCG alignments in clusters undergoing major mergers are no different from those in clusters in general. This indicates, at the least, that cluster mergers do not disrupt BCG alignments. Furthermore, any process by which mergers actually help build BCG alignments would have to act within the first $\sim$ 1 Gyr after first pericenter, because clusters in our sample are seen typically 1 Gyr after first pericenter and we see BCG alignments at full strength.  

According to \citet{Huang16} BCG-cluster alignments in general do exhibit trends with BCG axis ratio. Studying nearly 10,000 clusters, they found that alignment is weak to nonexistent for the roundest BCGs, strongest for moderately elliptical BCGs, and weaker again for the most elliptical BCGs.  However, \citet{Huang16} argue that this is not a physical trend. They found that their roundest BCGs had ill-defined position angles, and their most elongated BCGs were often galaxy mergers, where the PA was defined by local happenstance; each of these cases weaken what might be considered the baseline level of BCG-cluster alignment. We do not see the first trend because even our roundest BCGs have a well-determined PA due to an overwhelming signal-to-noise ratio. We do not see the second trend because of small sample size and because we discarded BCGs where the PA was ill-defined due to crowding.

In our merging systems, the projected separations are typically $\sim 1$ Mpc, and the subcluster separation vector is a proxy for the filament along which the subclusters presumably fell together.  Thus, our results provide strong evidence for a connection between the BCG scale and structures on scales of 1 Mpc or greater. 

\acknowledgments

This work was supported in part by NSF grant 1518246. Part of this was work performed under the auspices of the U.S. DOE by LLNL under Contract DE-AC52-07NA27344.  We thank Michael West for sharing the numbers behind the histogram in Figure 3 of \citet{West17BCG}. We also thank the anonymous referee for numerous constructive comments that improved this paper.

\facility{Subaru(SuprimeCam)}

\bibliography{ms}

\begin{thebibliography}{}
\expandafter\ifx\csname natexlab\endcsname\relax\def\natexlab#1{#1}\fi

\bibitem[{{Benson} {et~al.}(2017){Benson}, {Wittman}, {Golovich}, {Jee}, {van
  Weeren}, \& {Dawson}}]{BensonZwCl2341}
{Benson}, B., {Wittman}, D.~M., {Golovich}, N., {et~al.} 2017, \apj, 841, 7

\bibitem[{{Binggeli}(1982)}]{Binggeli82}
{Binggeli}, B. 1982, \aap, 107, 338

\bibitem[{{Catelan} \& {Theuns}(1996)}]{Catelan1996}
{Catelan}, P., \& {Theuns}, T. 1996, \mnras, 282, 455

\bibitem[{{Dawson}(2013)}]{Dawson2012}
{Dawson}, W.~A. 2013, \apj, 772, 131

\bibitem[{{Donahue} {et~al.}(2016){Donahue}, {Ettori}, {Rasia}, {Sayers},
  {Zitrin}, {Meneghetti}, {Voit}, {Golwala}, {Czakon}, {Yepes}, {Baldi},
  {Koekemoer}, \& {Postman}}]{Donahue16}
{Donahue}, M., {Ettori}, S., {Rasia}, E., {et~al.} 2016, \apj, 819, 36

\bibitem[{{Ensslin} {et~al.}(1998){Ensslin}, {Biermann}, {Klein}, \&
  {Kohle}}]{Ensslin1998}
{Ensslin}, T.~A., {Biermann}, P.~L., {Klein}, U., \& {Kohle}, S. 1998, \aap,
  332, 395

\bibitem[{{Feigelson} \& {Babu}(2012)}]{2012msma.book.....F}
{Feigelson}, E.~D., \& {Babu}, G.~J. 2012, {Modern Statistical Methods for
  Astronomy} (UK: Cambridge University Press)

\bibitem[{{Golovich} {et~al.}(2016){Golovich}, {Dawson}, {Wittman}, {Ogrean},
  {van Weeren}, \& {Bonafede}}]{GolovichMACS1149}
{Golovich}, N., {Dawson}, W.~A., {Wittman}, D., {et~al.} 2016, \apj, 831, 110

\bibitem[{{Golovich} {et~al.}(2017{\natexlab{a}}){Golovich}, {van Weeren},
  {Dawson}, {Jee}, \& {Wittman}}]{GolovichZwCl0008}
{Golovich}, N., {van Weeren}, R.~J., {Dawson}, W.~A., {Jee}, M.~J., \&
  {Wittman}, D. 2017{\natexlab{a}}, \apj, 838, 110

\bibitem[{{Golovich} {et~al.}(2017{\natexlab{b}}){Golovich}, {Dawson},
  {Wittman}, {Jee}, {Benson}, {Lemaux}, {van Weeren}, {Andrade-Santos},
  {Sobral}, {de Gasperin}, {Bruggen}, {Bradac}, {Finner}, \&
  {Peter}}]{Golovich18a}
{Golovich}, N., {Dawson}, W.~A., {Wittman}, D.~M., {et~al.} 2017{\natexlab{b}},
  ArXiv e-prints, arXiv:1711.01347

\bibitem[{{Golovich} {et~al.}(2018){Golovich}, {Dawson}, {Wittman}, {van
  Weeren}, {Andrade-Santos}, {Jee}, {Benson}, {de Gasperin}, {Venturi},
  {Bonafede}, {Sobral}, {Ogrean}, {Lemaux}, {Brada{\v c}}, {Br{\"u}ggen}, \&
  {Peter}}]{Golovich18b}
---. 2018, ArXiv e-prints, arXiv:1806.10619

\bibitem[{{Huang} {et~al.}(2016){Huang}, {Mandelbaum}, {Freeman}, {Chen},
  {Rozo}, {Rykoff}, \& {Baxter}}]{Huang16}
{Huang}, H.-J., {Mandelbaum}, R., {Freeman}, P.~E., {et~al.} 2016, \mnras, 463,
  222

\bibitem[{{Joachimi} {et~al.}(2015){Joachimi}, {Cacciato}, {Kitching},
  {Leonard}, {Mandelbaum}, {Sch{\"a}fer}, {Sif{\'o}n}, {Hoekstra}, {Kiessling},
  {Kirk}, \& {Rassat}}]{Joachimi15}
{Joachimi}, B., {Cacciato}, M., {Kitching}, T.~D., {et~al.} 2015, \ssr, 193, 1

\bibitem[{{Libeskind} {et~al.}(2013){Libeskind}, {Hoffman}, {Forero-Romero},
  {Gottl{\"o}ber}, {Knebe}, {Steinmetz}, \& {Klypin}}]{Libeskind2013}
{Libeskind}, N.~I., {Hoffman}, Y., {Forero-Romero}, J., {et~al.} 2013, \mnras,
  428, 2489

\bibitem[{{Niederste-Ostholt} {et~al.}(2010){Niederste-Ostholt}, {Strauss},
  {Dong}, {Koester}, \& {McKay}}]{Niederste-Ostholt10}
{Niederste-Ostholt}, M., {Strauss}, M.~A., {Dong}, F., {Koester}, B.~P., \&
  {McKay}, T.~A. 2010, \mnras, 405, 2023

\bibitem[{{Peng} {et~al.}(2010){Peng}, {Ho}, {Impey}, \& {Rix}}]{GALFIT2010}
{Peng}, C.~Y., {Ho}, L.~C., {Impey}, C.~D., \& {Rix}, H.-W. 2010, \aj, 139,
  2097

\bibitem[{{van Weeren} {et~al.}(2017){van Weeren}, {Andrade-Santos}, {Dawson},
  {Golovich}, {Lal}, {Kang}, {Ryu}, {Br{\`i}ggen}, {Ogrean}, {Forman}, {Jones},
  {Placco}, {Santucci}, {Wittman}, {Jee}, {Kraft}, {Sobral}, {Stroe}, \&
  {Fogarty}}]{vanWeeren2017Abell3411}
{van Weeren}, R.~J., {Andrade-Santos}, F., {Dawson}, W.~A., {et~al.} 2017,
  Nature Astronomy, 1, 0005

\bibitem[{{West}(1994)}]{West1994}
{West}, M.~J. 1994, \mnras, 268, 79

\bibitem[{{West} {et~al.}(2017){West}, {de Propris}, {Bremer}, \&
  {Phillipps}}]{West17BCG}
{West}, M.~J., {de Propris}, R., {Bremer}, M.~N., \& {Phillipps}, S. 2017,
  Nature Astronomy, 1, 0157

\end{thebibliography}

\end{document}